\documentclass{pasj00}
\begin{document}
\SetRunningHead{S. Karino}{Bimodality of Wind-fed Accretion}
%\Received{}%{yyyy/mm/dd}
%\Accepted{}%{yyyy/mm/dd}
%\Published{}%{yyyy/mm/dd}

\title{Bimodality of Wind-fed Accretion in High Mass X-ray Binaries}

%%% begin:list of authors
% Do NOT capitalize all letters in "textsc".
\author{Shigeyuki \textsc{Karino} %
%%  \thanks{Example: Present Address is xxxxxxxxxx}
}
\affil{
Faculty of Engineering, Kyushu Sangyo University, 2-3-1 Matsukadai, Higashiku, 
			 Fukuoka 813-8503 
}
\email{karino@ip.kyusan-u.ac.jp}

%%% end:list of authors

%%% Please use the following style in case that sorting by 
%%% affiliation is impossible. 
%
% \author{%
%   D-Firstname \textsc{D-Familyname}\altaffilmark{1}
%   E-Firstname \textsc{E-Familyname}\altaffilmark{1,2}
%   and
%   F-Firstname \textsc{F-Familyname}\altaffilmark{2}}
% \altaffiltext{1}{Address of Institute}
% \email{ddddd@xxx.xxx.xx.xx}
% \email{eeeee@xxx.xxx.xx.xx}
% \altaffiltext{2}{Address of Institute}

%% `\KeyWords{}' always has to be placed before `\maketitle'.
 %Do NOT move this preamble from here!

\maketitle

\begin{abstract}

%A neutron star fed by radiatively driven stellar wind from a supergiant primary can be a bright X-ray source.
%However, X-ray photoionization reduces the wind acceleration via line-driven mechanism. 

We study an influence of X-ray photo-ionization from an accreting neutron star in a high mass X-ray binary. 
%We investigate how the X-ray irradiation affects the wind flow and resulting accretion rates.
Our aim is to unveil a new principle governing X-ray luminosities of X-ray binaries, with a simple analysis of fluid equations simulating line-driven wind flow under influence of X-ray irradiation. 
%We adopt one dimensional accretion flow model including key physics. 
In this study, we solve equation of motion of the accretion flow taking into account the line-driven acceleration and X-ray photo-ionization. 
Under the influence of X-ray irradiation, we find the flow equations take two types of solutions. 
The first solution is characterized by a slow wind velocity which causes a large accretion rate. 
The second solution is a fast wind flow which results in a small accretion rate. 
We find that only the solution with a fast wind and faint X-ray luminosity is a steady solution.
On the other hand, slow wind solution with a large X-ray luminosity is not a realizable solution. 
In bright X-ray binary systems, X-ray luminosity would increase until strong X-ray reduces the line-driven acceleration and causes a stagnation of the wind. 
This result implies an important consequence that the X-ray luminosity of the wind-fed, X-ray emitting binary will be settled by the wind stagnation limit. 
At the same time, the fast wind solution with small X-ray luminosity also can be a steady state.  
Bright X-ray sources, such as Vela X-1, would have limiting luminosities of wind stagnation, while faint systems such a quiescent Super giant Fast X-ray Transients could be on the faint solutions.

\end{abstract}

\section{Introduction}

A high-mass X-ray binary (HMXB) is a close binary system consisting of a massive primary and a compact companion. 
In most cases, the X-ray emitting compact component is a neutron star (NS). 
HMXB systems are conventionally classified in two groups based on the optical component and temporal behaviors of the X-ray luminosity; Be-type and OB-type \citep{C84,C86}. 
In Be-type HMXB system, NS is usually in a wide elliptical orbit around a Be star. 
When the NS passes through the disk-like confined Be wind, it accretes plenty of gas, and a bright flare is observed in X-ray bands.
In contrast, in OB-type HMXB, a fraction of the stellar wind of the OB type super-giant (SG) star is captured by NS gravity, and produces X-rays as it falls onto the NS.
This OB-type HMXB has typically short orbital period and small eccentricity. 
%%Its X-ray luminosity increases up to $10^{36} \rm{erg s}^{-1}$. 

Winds of SG stars are driven by radiation absorbed in spectral lines \citep{CAK75}. 
Such stellar winds are usually referred to as line-driven wind. 
This line-driven mechanism (CAK mechanism) is enough effective to accelerate stellar wind up to several thousand $\rm{km \, s}^{-1}$. 
However, in X-ray binary systems, X-rays from NS significantly influence the SG stellar wind via X-ray photoionization of wind materials. 
The increase of highly charged ions decreases of the efficiency of the line-driven mechanism. 
Because of this effect, X-ray photoionization may significantly interfere with the wind acceleration. 
As a result, in HMXBs, wind velocities may become considerably slow \citep{SK90,S91,W06,KKS12}. 
The pioneering works on such an ionized-wind dynamics have been done by \citet{SK90} and \citet{S91}. 
They used a parameterized wind acceleration model and showed that the X-ray ionization changes the wind dynamics significantly. 
In the 2000s, 2D models of X-ray ionized wind have appeared. 
\citet{W06} used Monte Carlo simulations to obtain the ionization status around an accreting compact star, and compare their model with Vela X-1. 
\citet{KKS12} applied their statistical equilibrium technique to the line accelerated wind and obtained the 2D wind dynamics under the influences of X-ray ionization.

However, it is not easy to understand the association between wind velocity and X-ray luminosity in a HMXB. 
The degree of ionization can be low, when the X-ray luminosity of the NS is enough low. 
In this case, the velocity of the line-driven wind can be still large. 
Assuming Hoyle-Littleton accretion theory, a large wind velocity makes accretion rate smaller \citep{HL39,BH44}. 
If the accretion rate onto the compact component becomes smaller, the X-ray luminosity also becomes lower. 
As the consequence of this response of the wind velocity to the X-ray luminosity, it is theoretically possible to allow two types of accretion modes \citep{KKS12}. 
The first mode appears in the case of strong X-ray source which significantly affects the wind ionization state. 
This accretion mode produces a slow wind that can be accreted by the NS in a large amount. 
Consequently, the NS remains as a strong X-ray source. 
Hereafter, we call this type of accretion mode as a {\it{slow-bright}} mode. 
The second type of accretion modes may occur in the case of a weak X-ray source that does not significantly influence the wind ionization state. 
This second mode results in faster wind solution. 
We call this accretion mode as a {\it{fast-faint}} mode.

If it is confirmed that the bimodality of the accretion flow really exists, it contributes to the promotion of understanding X-ray properties of HMXBs. 
For example, it may help to understand the observed difference of wind-fed binaries, such as normal HMXBs, Super giant Fast X-ray Transients (SFXTs), and recently observed faint X-ray sources \citep{WIR06, L13}. 
Furthermore, if some perturbations cause switching between two accretion modes in particular system, a drastic variability of X-ray luminosity may be produced \citep{KKS12}.

The bimodality of the accretion mode is suggested by \citet{KKS12}, however, it has not been verified in detail. 
Hence, in this study, we show that the wind equations describing HMXB winds have two types of solutions. 
To show this, we adopt one dimensional accretion flow model. 
In all the previous studies, the X-ray luminosity of the compact star, $L_{\rm{X}}$, has been treated as a parameter. 
To set this parameter, they referred the X-ray luminosity of the observed bright HMXB, such as Vela X-1. 
However, unless we relate the dynamics of accretion flow to the resultant X-ray luminosity, we never verify the bimodality of the wind flow suggested by \citet{KKS12}. 
Therefore, in our computations, the X-ray luminosity of the neutron star is self-consistently solved at the same time as the wind dynamics such as density, velocity and so on.  
Additionally, we compare our results with the observed HMXB systems, and we point out that observed HMXBs are stuck in the steady states in the parameter space.

In the next section, we introduce our accretion model. 
We show the method to solve the fluid equations duplicating line-driven wind under the influence of the X-ray photoionization. 
In section 3, we show our numerical results. 
Section 4 is devoted to discussions. 
We compare our solutions with the observed HMXB systems. 
Additionally, we discuss the key factor to govern the luminosity of wind-fed X-ray binaries. 
A short summary is given in the last section.

%%%%%%%%%%%%%%%%%%%%%%%%%%%%%%%%%%%%%%%%%%%%%%%%%%

\section{Numerical Model}

To reproduce a SG wind which is accreted onto a NS, we adopt one dimensional model used in \citet{SK90} and \citet{S91}. 
In this model, our calculations are restricted only to the line of centers of the binary components. 
That is, we solve a wind flow that linearly connects a SG surface to a NS.
We do not consider flow motion deviates from this straight track. 
The equations to be solved in this model are the mass conservation low
\begin{equation}
4 \pi r^2 \rho v = \dot{M} ,
\label{eq:1}
\end{equation}
and the equation of motion
\begin{equation}
v \frac{d v}{d r} = - \frac{1}{\rho} \frac{d \phi}{dr} - \frac{1}{\rho} \frac{dp}{dr} + g_{\rm{R}} .
\label{eq:2}
\end{equation}
$r$, $\rho$, $v$, and $p$ denote the distance from the center of the SG, wind density, wind velocity and pressure, respectively. 
We use the symbol $r$ as the distance measured from the center of the SG hereafter, while we use $r_{\rm{X}}$ as the distance from the NS. 
In order to relate the density and pressure, we assume $p = K_{\rm{pol}} \rho^{\gamma}$, with $\gamma = 5/3$. 
Here, $K_{\rm{pol}}$ is a constant and we obtain this value from the surface density and pressure at the SG surface (to be shown later). 
$\dot{M}$ is the mass loss rate of the SG via stellar wind. 
$\phi$ is the gravitational potential due to both SG and NS. 
The last term $g_{\rm{R}}$ is the sum of the contributions of radiation forces. 
This term includes the radiation pressure terms due to SG and NS. 
The contribution from the SG consists of radiation pressure and line-driven acceleration. 
To obtain radiation forces, we use almost the same procedure used in \citet{N13} whose aim is to compute the line-driven disk wind of quasars. 
Using UV and X-ray fluxes, $F_{\rm{UV}}$ and $F_{\rm{X}}$, the last term in Eq.~(\ref{eq:2}) can be written as 
\begin{equation}
g_{\rm{R}} = (1+M_{\rm{f}}) \frac{\sigma_{\rm{e}} F_{\rm{UV}} }{c} 
+ \frac{\sigma_{\rm{X}} F_{\rm{X}} }{c} .
\end{equation}
Here, $\sigma_{\rm{e}}$ and $\sigma_{\rm{X}}$ are electron (Thomson) scattering coefficients.  
They are simply related as $\sigma_{\rm{X}} = 100 \sigma_{\rm{e}}$ unless the degree of ionization is extremely high \citep{N13}.

The efficiency of the line-acceleration is included via force multiplier, $M_{\rm{f}}$ \citep{CAK75}. 
Here, we use the modified version of the force multiplier \citep{OCR88}, and its parameterizations given by \citet{SK90}; 
\begin{equation}
M_{\rm{f}} = k t^{-0.6} \left[ \frac{ (1+t \eta)^{0.4} - 1 }{ (t \eta)^{0.4} } \right] .
\end{equation}
In this equation, $k$ and $\eta$ are parameters introduced by \citet{SK90}, and they are calculated as
\begin{equation}
k = 0.03 + 0.385 \rm{exp} (-1.4 \xi^{0.6} )
\end{equation}
and 
\begin{eqnarray}
\log \eta = \left\{ \begin{array}{ll}
6.9 \exp \left( 0.16 \xi^{0.4} \right) & ( \log \xi \le 0.5 ) \\
9.1 \exp \left( -7.96 \times 10^{-3} \xi \right) & ( \log \xi > 0.5 ) 
\end{array} \right.
,
\end{eqnarray}
respectively. 
$t$ is a parameter called a local optical depth, and it is given as
\begin{equation}
t = \sigma_{\rm{e}} \rho v_{\rm{th}} \left| \frac{dv}{dr} \right|^{-1} . 
\end{equation}
The ionization factor $\xi$ is defined as 
\begin{equation}
\xi = \frac{m_{\rm{p}} L_{\rm{X}} }{\rho r_{\rm{X}}^2} e^{- \tau_{\rm{X}}} , 
\end{equation}
where $m_{\rm{p}}$ is a proton mass, and $r_{\rm{X}}$ is a distance from the NS. 
$\tau_{\rm{X}}$ is the optical depth related to X-ray, which is measured from the NS position.

In the computation of the accretion flow, we need to obtain the optical depth at each point. 
In one dimensional computation, the optical depths are given as
\begin{equation}
%%\tau_{\rm{UV}} = \sum_{j} \rho \sigma_{\rm{e}} \Delta s
%%\tau_{\rm{UV}} ( r ) = \int_{R_{\rm{SG}}}^{r} \rho \sigma_{\rm{e}} ds
\tau_{\rm{UV} , \it{j}} = \sum_{i=1}^{j} \rho_{i} \sigma_{\rm{e} } \Delta s
\end{equation}
and 
\begin{equation}
%%\tau_{\rm{X}} = \sum_{j} \rho \sigma_{\rm{X}} \Delta s,
%%\tau_{\rm{X}} ( r_{\rm{X}} ) = \int_{0}^{r_{\rm{X}}} \rho \sigma_{\rm{X}} ds
\tau_{\rm{X} , \it{j}} = \sum_{i=j}^{N} \rho_{i} \sigma_{\rm{X} } \Delta s
\end{equation}
where the subscript $j$ means the values at the $j$-th computational grid, numbered from the SG surface. 
$\tau_{\rm{UV}}$ is an optical depth for UV radiation measured from the SG, and $\tau_{\rm{X}}$ is that for X-ray measured from the NS, respectively. 
$N$ is the total grid number, and here we use 4096 grids to compute the wind flow from the SG surface, $R_{\rm{SG}}$, to the NS position, $R_{\rm{NS}}$. 
$\Delta s = (R_{\rm{NS}} - R_{\rm{SG}})/4096$ is the uniformly-split grid size. 
With these optical depths at each point, fluxes are related to the luminosities by 
\begin{equation}
F_{\rm{UV}} = \frac{ L_{\rm{UV}} }{ 4 \pi r^2 } e^{- \tau_{\rm{UV}} } 
\end{equation}
and 
\begin{equation}
F_{\rm{X}} = \frac{ L_{\rm{X}} }{ 4 \pi r_{\rm{X}}^2 } e^{- \tau_{\rm{X}} } .
\end{equation}
To obtain the optical depths, it is required iterative computations.

\begin{figure}
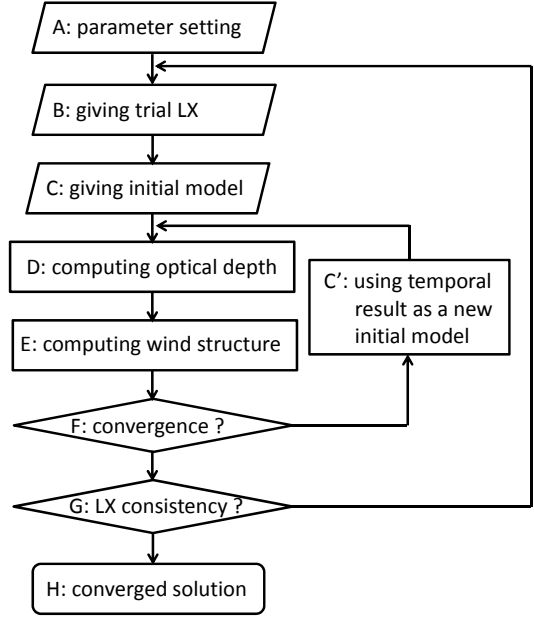

  \begin{center}
    \FigureFile(70mm,90mm){re_fig_1.eps}
  \end{center}
  \caption{
	Flowchart of the computation.
  }
  \label{fig:flowchart}
\end{figure}

For the iterative computations, firstly we give a provisional density profile of the accretion flow (trial density). 
Then we compute the optical depth everywhere along the flow.
Using this optical depth distribution, we solve equations (1) and (2) from the SG surface toward the NS. 
To integrate the equation of motion, we use the standard Runge-Kutta scheme. 
The integration is started from the SG surface, $r = R_{\rm{SG}}$. 
At this position, we assume the initial conditions as the following. 
As the initial condition, we assume a slow wind velocity $v_{\rm{w,0}} = 5.0 \times 10^{6} \rm{cm \, s}^{-1}$. 
With this value, the initial density is calculated by Eq.~(\ref{eq:1}) as $\rho_0 = \dot{M} / (4 \pi R_{\rm{SG}}^{2} v_{\rm{w,0}})$. 
The temperature is assumed to be $T_0 = 40 \, 000 \rm{K}$ and it leads the initial pressure $p_0 = (2 k_{\rm{B}} / H) \rho_0 T_0$. 
$k_{\rm{B}}$ is the Boltzmann constant and $H$ denotes the atomic mass unit. 
During the integration, we check the balance between the potential energy due to NS gravity and the kinetic energy of the wind matter. 
If the NS potential overcomes the kinetic energy, we stop the computation. 
This end point corresponds to the accretion radius in Bondi-Hoyle-Littleton theory; 
\begin{equation}
r_{\rm{acc}} = \frac{2GM_{\rm{NS}}}{v_{\rm{w}}^2} 
\label{eq:racc}
\end{equation}
\citep{HL39,BH44}.
%This accretion radius is the distance measured from the NS, not the SG center. 
Once we finish the computation, we can obtain the density and velocity function along the flow. 
Then we compute the derivation of the obtained solution form the initial density profile. 
If the deviation is large, we set the obtained solution as a new trial density. 
We repeat these procedures until the deviation becomes enough small ($\Sigma \left| d \rho / \rho \right| < 0.01$). 
The above scheme corresponds to Part C to Part F in Fig.~\ref{fig:flowchart}, which shows the flowchart of the whole computation.

Additionally, in order to obtain the consistent solution, the X-ray luminosity from the NS is required. 
However, it cannot be obtained unless the accretion rate is known.
Hence, before the iterative procedure introduced above, we give a provisional X-ray luminosity (trial $L_{\rm{X}}$). 
With this trial $L_{\rm{X}}$, we solve the flow equations (\ref{eq:1}) and (\ref{eq:2}) iteratively. 
Then, from the obtained solution, we compute the accretion rate onto the NS. 
To this end, we check the accretion radius from equation (\ref{eq:racc}), and calculate the accretion rate by
\begin{equation}
\dot{m} = \pi r_{\rm{acc}}^{2} \rho_{\rm{acc}} v_{\rm{acc}} .
\label{eq:dotm}
\end{equation}
Here, $\rho_{\rm{acc}}$ and $v_{\rm{acc}}$ denote the density and the velocity at the accretion radius.
With this accretion rate, the X-ray luminosity is obtained as $L_{\rm{X}} = \eta \dot{m} c^2$. 
$\eta$ denotes the efficiency of X-ray emission, and it is assumed to be 0.025 in this work. 
If the deviation between this obtained X-ray luminosity and the trial $L_{\rm{X}}$ is enough small  ($ \left| d L_{\rm{X}} / L_{\rm{X}} \right| < 0.01$), we adopt this obtained wind solution as a {\it{converged solution}}. 
This second level iteration is shown in Fig.~\ref{fig:flowchart}, Part B to G. 

In this study, we sweep the UV luminosity of the SG, meanwhile basically we fix other parameters. 
With these procedures, we investigate the location of the converged solutions in the parameter space. 
Later on, we change the parameters of SG and NS, and compare the results.

%%%%%%%%%%%%%%%%%%%%%%%%%%%%%%%%%%%%%%%%%%%%%%%%%%

\section{Result}

\begin{figure}
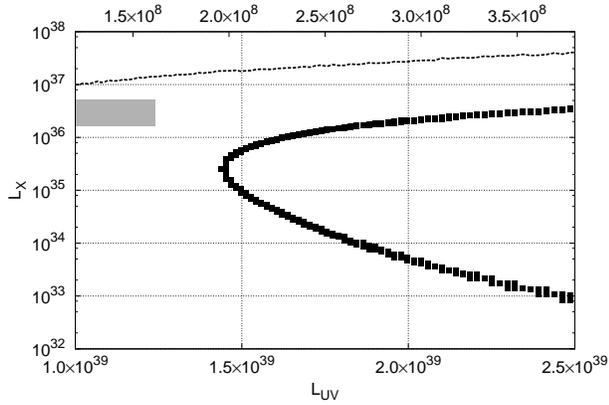

  \begin{center}
    \FigureFile(80mm,80mm){re_fig_2.eps}
  \end{center}
  \caption{
Locations of converged solutions in $L_{\rm{UV}} - L_{\rm{X}}$ plane.
The converged solutions are indicated by filled squares. 
We can see clear two branches corresponding to the slow-bright sequence and the fast-faint sequence. 
The parameter set is tuned to simulate Vela X-1 system (see Section 3). 
The upper dashed line denotes the stagnation limit where wind velocity becomes zero somewhere along the flow. 
%Asterisk indicates the position of Vela X-1 in this parameter space. 
Square shaded region indicates the position of Vela X-1 in this parameter space. 
The scales at the top of figure indicate the wind velocity corresponds to the horizontal axis. 
  }
  \label{fig:main}
\end{figure}

First, we show our result which is obtained with parameter set $\dot{M} = 5.0 \times 10^{-6} M_{\odot}$, $M_{\rm{SG}} = 23.5 M_{\odot}$, $R_{\rm{SG}} = 30 R_{\odot}$, $M_{\rm{NS}} = 1.88 M_{\odot}$ and $P_{\rm{orb}} = 8.9 \rm{d}$. 
This parameter set is chosen to be in tune with the observed parameters of the prototypical wind-fed HMXB, Vela X-1 \citep{vK95,W06}. 
We sweep $L_{\rm{X}}$ - $L_{\rm{UV}}$ space to seek the {\it{converged solution}} which has been introduced in the last section. 
The result of the numerical computation is shown in Fig.~\ref{fig:main}. 
In this figure, the locations of converged solutions in the parameter space are indicated with filled squares. 
The vertical axis denotes the converged X-ray luminosity from the NS. 
The horizontal axis denotes the UV luminosity of the SG.
At the top of the figure, the velocity of the stellar wind is also shown. 
This wind velocity is computed by Eqs.~(\ref{eq:1}) and (\ref{eq:2}) at $2 R_{\rm{SG}}$ %as follows. 
%For given UV luminosity, we compute wind solution 
without all effect of the NS (gravity, X-ray emission, and ionization). 
The wind velocity profile that obtained this computation can be well fitted with the wind velocity formula in CAK theory; 
\begin{equation}
v_{\rm{w}} = v_{\infty} \left[ 1 - \frac{R_{\rm{SG}}}{r} \right]^{\beta} .
\end{equation}
%In Fig.~Z, we show in example of the wind velocity profile. 
%Then we adopt the wind velocity at $r = 20 M_{\rm{SG}}$. 
%The corresponding $L_{\rm{UV}}$ is also shown by the scales at the top of Fig.~\ref{fig:main}.

When the X-ray radiation from the NS is very strong, the degree of ionization becomes high. 
In this case, the CAK process cannot be effective and the wind cannot be accelerated enough to overcome the SG gravity. 
That is, the wind would be {\it stagnated} \citep{KKS12}. 
The dashed line in the upper region of Fig.~\ref{fig:main} denotes the critical X-ray luminosity to cause this wind stagnation. 
Above this critical $L_{\rm{X}}$, the wind velocity becomes zero somewhere along the flow.

\begin{figure}
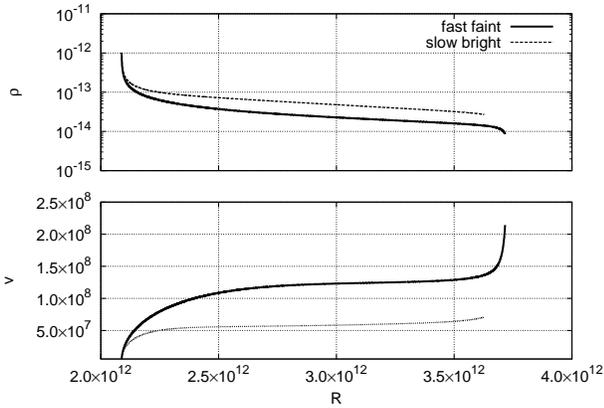

  \begin{center}
    \FigureFile(80mm,80mm){re_fig_3.eps}
  \end{center}
  \caption{
Density (upper panel) and velocity (lower panel) profile of the selected solutions. 
The solid and dashed lines show the fast-faint solution and the slow-bright solution, respectively. 
These solutions correspond to the two sequence at $L_{\rm{UV}} = 1.75 \times 10^{39} \rm{erg \, s^{-1}}$ in Fig.~\ref{fig:main}. 
  }
  \label{fig:profile}
\end{figure}

From the distribution of the converged solutions in Fig.~\ref{fig:main}, we can see that there are no converged solution in low UV luminosity ($=$slow wind) region ($\le 1.4 \times 10^{39} \rm{erg \, s^{-1}}$). 
It means that the X-ray luminosity does not have steady state and varies with time, when the wind velocity is slow (= UV luminosity is low). 
A band of the converged solutions first appears as a band at $L_{\rm{UV}} \approx 1.4 \times 10^{39} \rm{erg \, s^{-1}}$. 
With increasing wind velocity, the converged solutions make clear two sequences. 
Upper sequence is X-ray bright solutions; $L_{\rm{X}} > 10^{36} \rm{erg \, s^{-1}}$. 
%This solution sequence corresponds to the slow mode accreting state. 
%The converged wind profile belonging to this bright sequence has a slow velocity and a large density distribution. 
On the other hand, the lower sequence has low-X-ray luminosities; $L_X < 10^{34} {\rm erg \, s^{-1}}$. 
%This sequence corresponds to the fast mode accretion. 
%Wind solution of this sequence has a fast wind and a low density profile. 
In Fig.~\ref{fig:profile}, we show the density and velocity profiles of the converged solutions, with $L_{\rm{UV}} = 1.75 \times 10^{39} {\rm erg \, s^{-1}}$. 
The solid lines in these figures show the profiles of wind that belongs to the lower (faint) sequence. 
These solid lines clearly show that the flow of this branch has lower density and fast wind velocity. 
We call this sequence of solutions as {\it{fast-faint}} solutions. 
In this solution, the wind velocity rapidly increases near the NS. 
This is because near the NS, the gravity force of NS strongly accelerates the wind flow. 
The gain of wind velocity causes two consequences; decrease of the density and shrink of the accretion radius. 
These effects lead a decay of mass accretion rate onto the NS. 
Especially, the shrinkage of the accretion radius due to the large velocity is a dominant effect on the decrease of mass accretion rate (see Eq.~(\ref{eq:dotm}), also see Eq.~(\ref{eq:racc})).
On the other hand, dashed lines show the upper (bright) branch solution and it has high density and low velocity. 
We call this sequence of solutions as {\it{slow-bright}} solutions. 
Since this solution derives a large accretion radius, the wind profile ends at a small distance from the SG. 
Therefore, in this solution the acceleration via NS gravity cannot work before the flow passes accretion radius. 
Although the flow structure inside the accretion radius is quite complicated, a bow shock would be formed near the accretion radius and flow matter would be captured by NS magnetosphere.

Such an existence of two branches demonstrates just the bimodality of the accretion flow suggested by \citet{KKS12} in a tangible way.
However, among these two solutions, only the fast-faint solution is a viable solution (see Sec. 4). 
The slow-bright sequence is unstable solution, and actual bright systems should concentrate not this slow-bright sequence but the stagnation limit.

\begin{figure}
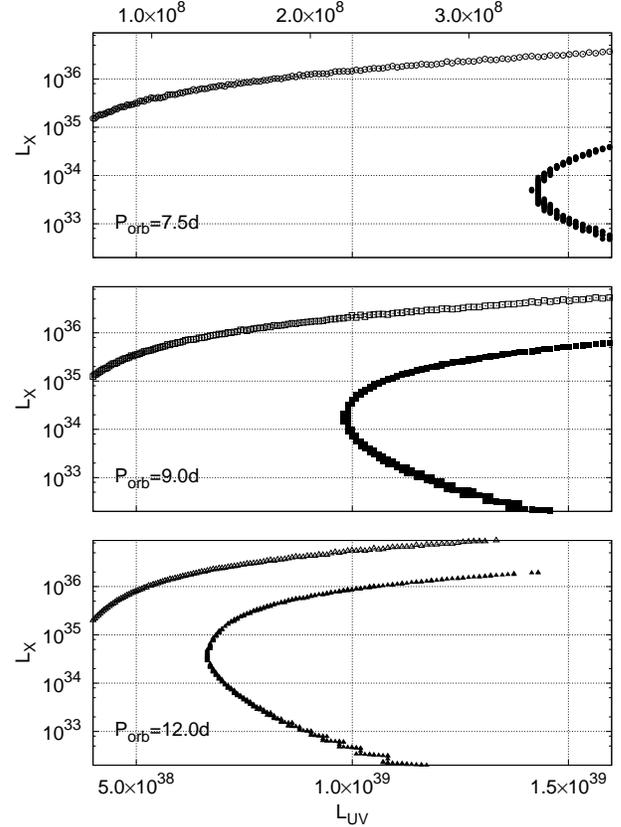

  \begin{center}
    \FigureFile(80mm,80mm){re_fig_4.eps}
  \end{center}
  \caption{
The locations of converged solutions in $L_{\rm{UV}} - L_{\rm{X}}$ plane. 
We vary the orbital period of the system: $P_{\rm{orb}} = 7.5 \rm{d}$ (upper pannel), 9 d (middle) and 12 d (lower). 
The stagnation limits are also shown with open symbols. 
The scales at the top of figure indicate the wind velocity corresponds to the horizontal axis. 
  }
  \label{fig:porb}
\end{figure}

\begin{figure}
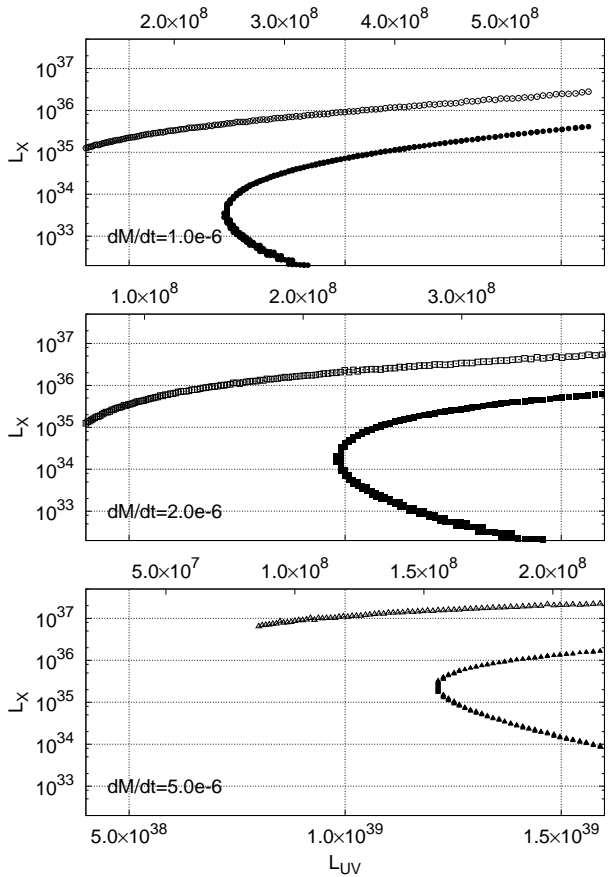

\begin{center}
\FigureFile(80mm,80mm){re_fig_5.eps}
\end{center}
\caption{
The locations of converged solutions in $L_{\rm{UV}} - L_{\rm{X}}$ plane. 
We vary the wind mass loss rate from SG: $\dot{M} = 1 \times 10^{-6} M_{\odot} \rm{yr}^{-1}$ (upper pannel), $2 \times 10^{-6} M_{\odot} \rm{yr}^{-1}$ (middle) and $5 \times 10^{-6} M_{\odot} \rm{yr}^{-1}$ (lower). 
The scales at the top of figures indicate the wind velocity corresponds to the horizontal axis. 

}
\label{fig:mdot}
\end{figure}

Results obtained with different parameter sets are shown in Figs.~\ref{fig:porb} and \ref{fig:mdot}. 
In these figures, we alter the NS mass to $M_{\rm{NS}} = 1.4 M_{\odot}$, while the SG mass and radius are the same with Fig.~\ref{fig:main}. 
In Fig.~\ref{fig:porb}, we show the results when we vary the orbital period of the system; $P_{\rm{orb}} = 7.5 \rm{d}$ (upper panel), 9d (middle panel) and 12d (lower panel). 
In this case, the mass loss rate is fixed as $\dot{M} = 2.0 \times 10^{-6} M_{\odot} \rm{yr}^{-1}$. 
The location of the solutions is strongly depends on the orbital period. 
When the orbital period is short, we cannot obtain the converged solutions in almost searched region. 
However, the stagnation limit is almost unchanged as $L_{\rm{X}} \approx 10^{36} \rm{erg \, s^{-1}}$ in this case. 
In Fig.~\ref{fig:mdot}, we show the results when we vary the mass loss rate of the SG, while the orbital period is fixed as $P_{\rm{orb}} = 9 \rm{d}$. 
We compute three cases with $\dot{M} = 1 \times 10^{-6} M_{\odot} \rm{yr}^{-1}$ (upper panel), $2 \times 10^{-6} M_{\odot} \rm{yr}^{-1}$ (middle panel) and $5 \times 10^{-6} M_{\odot} \rm{yr}^{-1}$ (lower panel). 
%%The minimum $L_{\rm{UV}}$ where the solution exists are $3.5 \times 10^{38}$, $4.6 \times 10^{38}$ and $5.9 \times 10^{38} \rm{erg \, s}^{-1}$ for these three case in the same sequence. 
The location of the solutions is also depends on the mass loss rate via stellar wind. 
When the mass loss rate is small, the branches of solutions locate in a high velocity region. 
On the other hand, when the mass loss rate is large, the solutions shift toward a low velocity region, while it corresponds to high UV luminosity region. 
This time, the stagnation luminosity shifts upward with the mass loss rate from the SG.

%%%%%%%%%%%%%%%%%%%%%%%%%%%%%%%%%%%%%%%%%%%%%%%%%%

\section{Discussion}

%\subsection{What settles X-ray luminosities} 

\begin{figure}
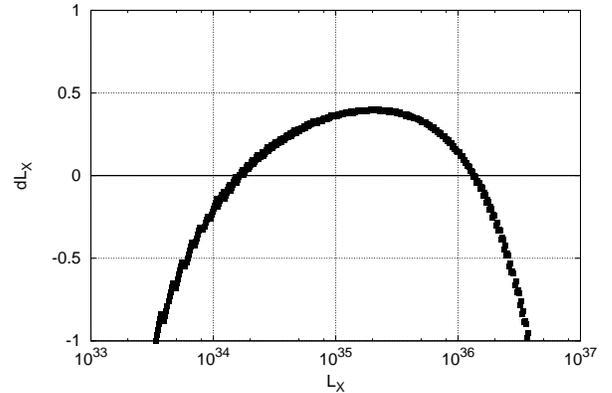

  \begin{center}
    \FigureFile(80mm,80mm){re_fig_6.eps}
    %%% \FigureFile(width,height){filename}
  \end{center}
  \caption{
Difference between the trial $L_{\rm{X}}$ and the obtained X-ray luminosity, $d L_{\rm{X}}$
In this case the UV luminosity is chosen to be $1.75 \times 10^{39} \rm{erg \, s^{-1}}$ and other parameters are in accordance with Fig.~\ref{fig:main}.
We obtain converged solutions when $dL_{\rm{X}} = 0$. 
}
\label{fig:difLX}
\end{figure}

In order to obtain the converged solution sequences, we sweep the parameter space constituted by X-ray luminosity of the NS and UV luminosity of the SG (velocity of the stellar wind). 
In Fig.~\ref{fig:difLX}, difference between the trial $L_{\rm{X}}$ and the obtained X-ray luminosity, $d L_{\rm{X}} = L_{\rm{X, Trial}} - L_{\rm{X, Obtained}}$, is shown. 
We obtain the converged solutions when $d L_{\rm{X}} = 0$. 
In this figure, we fix the UV luminosity as $L_{\rm{UV}} = 1.75 \times 10^{39} \rm{erg \, s^{-1}}$ (or $v_{\infty} = 2.5 \times 10^{8} \rm{cm \, s}^{-1}$) and vary the trial $L_{\rm{X}}$.
Other parameters are the same with Fig.~\ref{fig:main}. 
Between two zeros (two solutions), the difference $dL_{\rm{X}}$ becomes positive. 
That is, in this region, the emerged $L_{\rm{X}}$ is smaller than the trial $L_{\rm{X}}$. 
Since $L_{\rm{X}}$ is proportional to $\dot{m}$, it means that the resultant mass accretion rate is smaller than the initial prediction. 
Hence, systems located between two sequences cannot gain its X-ray luminosity. 
Rather, systems located in this region would become fainter. 
In $L_{\rm{UV}} - L_{\rm{X}}$ plane, it means that the systems between two sequences would shift downward with time. 
On the other hand, in other regions, the difference $d L_{\rm{X}}$ is negative. 
Namely, the resultant X-ray luminosity overcomes the expected (trial) X-ray emissivity. 
It means that the system in this region grows its X-ray luminosity gradually. 
Hence systems outside of the mid-band region between two sequences will shift upward in $L_{\rm{UV}} - L_{\rm{X}}$ plane.
Consequently, the solutions along the fast-faint sequence are {\it{steady state}}. 
Binary systems on this sequence keep faint status, unless drastic change of the wind status occurs. 
On the other hand, the solutions along the slow-bright sequence are {\it{unsteady state}}. 
Rather, this sequence is a dividing ridge of the evolutionary passes of HMXB systems. 
In $L_{\rm{UV}} - L_{\rm{X}}$ plane, systems located below this sequence will decrease their X-ray luminosities until arriving at the fast-faint sequence. 
Meanwhile, systems located above this sequence will increase their X-ray luminosities until arriving at the stagnation limit. 
Systems coming at the stagnation limit cannot be treated in the present framework. 
The accretion flow at this limit will be governed by complex combinations of pressure, radiation, system rotation, and so on \citep{BKFT90}.  
Since such a state should be highly non-linear and unsteady, it is out of the present scope of work. 
However, it is quite possible that systems as bright as the stagnation limit will stay just below this limiting luminosity. 
Because, if the X-ray luminosity overcomes this limit, the mass supply from the accretion flow will stop. 
Then, the system will be pushed back below the limiting X-ray luminosity. 
Hence, brighter luminosity above the stagnation limit would be {\it{forbidden}} \citep{KKS12}. 
Systems which have small terminal velocities cannot take any converged solution. 
Hence, these systems would inevitably concentrate just below the stagnation limit.

This result implies an important consequence. 
That is, the X-ray luminosity of the wind-fed X-ray binary will be settled either by the stagnation limit, or by the fast-faint sequence of the converged solutions. 
These settled luminosities are much smaller than Eddington luminosity. 
The Eddington luminosity for a NS is $\sim 10^{38} \rm{erg \, s^{-1}}$ and it is higher than the stagnation limit obtained in this work and previous works \citep{KKS12}. 
It suggests that the wind-fed systems cannot achieve the Eddington luminosity.
Because, before that the mass supply via stellar wind will be stopped due to the X-ray induced stagnation. 
However, one dimensional treatment may be insufficient to understand the true nature in rotating systems such as HMXBs. 
Further studies are required in multi-dimensional numerical approaches \citep{BKFT90}.

%\subsection{Observed systems}

Next, we compare our results with some observed X-ray luminosities of HMXBs. 
In general, wind-fed X-ray binaries with SG have orbital period around 10 days \citep{C86}. 
%Typical mass loss rate via stellar wind is $10^{-6} M_{\odot} \rm{yr}^{-1}$. 
Among them, Vela X-1 is a prototypical wind-fed X-ray binary system that has long history of the observations. 
Since Vela X-1 is an eclipsing binary, its orbital parameters have been well studied \citep{vK95,W06}. 
%%In a long history of observations, detailed binary parameters have been known; 
%%$\dot{M} = 5 \times 10^{-6} M_{\odot} \rm{yr}^{-1}$，$R = 30 R_{\odot}$，$M = 23.5 M_{\odot}$，$P_{\rm{orb}} = 8.9 \rm{d}$，$m = 1.88 M_{\odot}$. 
Adopting parameters introduced in Sec.~3, we obtain the converged solution sequences and we show it in Fig.~\ref{fig:main}. 
%In this figure, Vela X-1 is shown by an asterisk. 
In this figure, Vela X-1 locates in the square shaded region which shows uncertainty of the observations.
According to observations, the wind velocity should be in the range between $1.1 \times 10^{8} \rm{cm \, s^{-1}}$ \citep{W06} and $1.6 \times 10^{8} \rm{cm \, s^{-1}}$ \citep{DSM09}. 
The X-ray luminosity is varying, but roughly in the range of $1 - 5 \times 10^{36} \rm{erg \, s^{-1}}$ \citep{W06,FKP10}. 
It is clear that the location of Vela X-1 in this figure is just below of the stagnation limit. 
This is previously pointed by \citet{KKS12}, and it is also consistent with our previous discussions. 
If the observed wind terminal velocity ($1.1 - 1.6 \times 10^{8} \rm{cm \, s^{-1}}$) is correct, Vela X-1 has no converged solution since its wind velocity is too slow. 
Hence, this system cannot take a faint phase and will keep being bright X-ray source, unless the mass donor evolves any further. 
As another example, 4U 2206+54 is a HMXB system with $P_{\rm{orb}} = 9.6 \rm{d}$ and this orbital period is similar to Velar X-1. 
In this system, the stellar wind from supergiant primary is estimated as $\sim 3.5 \times 10^{7} \rm{cm \, s^{-1}}$. 
This wind velocity is quite slow and the origin of this extremely slow wind is one of remained puzzles \citep{R06}. 
Since this system has a large orbital eccentricity, our model cannot be applied directly. 
However, we consider that this slow wind velocity may be another example of the stagnated wind due to X-ray irradiation.

Besides bright sources, recently, fainter persistent sources (with $L_{\rm{X}}$ less than $10^{35} \rm{erg \, s^{-1}}$) have been found by monitoring observations \citep{WIR06, L13}. 
Especially, in Galactic central region, there are many faint X-ray sources.
Among them, wind-fed systems take a substantial portion \citep{PRP02}. 
Furthermore, some SFXTs in quiescent phase can be thought as faint persistent sources \citep{WIR06}. 
In quiescent phase, SFXTs show very faint X-ray luminosity, much less than classical wind-fed HMXBs \citep{SBL08}. 
These faint HMXBs and quiescent SFXTs may locate along the fast-faint sequence of our converged solutions. 
The X-ray luminosity of the fast-faint sequence depends on the UV-luminosity (or equivalently the wind velocity), orbital period, mass loss rate of SG, and so on (see Figs.~\ref{fig:porb} and \ref{fig:mdot}).
In any case, the fast faint sequence strides across X-ray luminosity band of $10^{32}$ - $10^{35} \rm{erg/s}$, that is typically observed in faint X-ray sources. 
Furthermore, the SG stars in SFXT systems have large mass loss rates ($\dot{M} = 10^{-(5-6)} M_{\odot} \rm{yr}^{-1}$) \citep{WZ07}. 
This is consistent with our results shown in Figs.~\ref{fig:mdot}.
Namely, with a large mass loss rate, steady fast-faint sequence appears even if the wind velocity is moderate.  
In the case of SFXT, the system increases its X-ray luminosity due to some reasons such as clumpy wind accretion \citep{I05,WZ07}, accretion from disk-like wind flow of Be star \citep{SRM07,RSM07}, magnetic gating effect \citep{BFS08}, and so on. 
However, if an X-ray flare luminosity of a SFXT is less than $10^{36-37} \rm{erg \, s^{-1}}$, the system remains still under the dividing ridge given by a slow-bright sequence. 
Hence, after a flare, the system comes back to the faint state immediately.

Since SFXTs and faint HMXBs are newly recognized objects, complete parameter sets have not been provided. 
Additionally, since SFXTs have large eccentricities \citep{K10}, one dimensional approach may be restricted. 
However, it has been shown that some SG stars hosting SFXTs provide very large wind velocity. 
When the wind velocity is very large, it dominates the system orbital velocity. 
In this case, therefore, our one dimensional model would have high validity. 
In the future, when the observed data of faint sources increase, it may be verified that these faint objects actually locate on our fast-faint sequences.

Of course, such 1D treatment of the accretion flow is not sufficient to apply the actual binary systems. 
Especially, in short $P_{\rm{orb}}$ systems, the orbital motion can never be neglected. 
Because of the non-radial motion of the wind matter, in 2D/3D cases, the matter can accretes also from the side of the main radial flow-line. 
Such sideways accretion would make a large influence on the accretion rate.
Especially, the stagnation limit of the flow will change its meaning. 
When the NS has an orbital motion, even if the wind flow stagnates, the NS would have a mass supply from the sideways accretion. 
In this case the accretion radius and the subsequent accretion rate will be dominated by the orbital velocity; $r_{\rm{acc}} = 2GM_{\rm{NS}} / v_{\rm{orb}}^2$.   
The orbital velocity can be calculated with 
\begin{equation}
v_{\rm{orb}} = \sqrt{ \frac{M_{\rm{SG}}G}{r_{\rm{orb}}} }
\end{equation}
and
\begin{equation}
v_{\rm{orb}} = \frac{2 \pi r_{\rm{orb}}} { P_{\rm{orb}}}
\end{equation}
when the NS mass is much smaller than the SG mass. 
Assuming $P_{\rm{orb}} = 10 \rm{d}$, the orbital velocity becomes $2.8 \times 10^{7} \rm{cm \, s^{-1}}$. 
This value is smaller than the typical slow-wind velocity ($\sim 5 \times 10^{7} \rm{cm \, s^{-1}}$, see Fig.~\ref{fig:profile}), then the 1D approximation could have a meaning in our main models. 
However, for systems with short orbital periods, and situations around stagnation limits, multidimensional, time-dependent approach is obviously required. 
The asymmetry of X-ray radiation from the NS (for instance, radiation from a disk) could be a problem. 
Although 1D model may be too much simplified, 1D treatment is the fastest way to numerically verify the bimodality of the accretion flow in HMXB suggested by \citet{KKS12}.
Now, it has been done. 
We consider that this work motivates further theoretical and observational studies of such a bimodality and it consequences to the X-ray luminosities of HMXBs.

%%%%%%%%%%%%%%%%%%%%%%%%%%%%%%%%%%%%%%%%%%%%%%%%%%

\section{Conclusion}

In this paper, we studied one dimensional model of wind-fed accretion flow taking into account X-ray photo ionization due to an accreting neutron star. 
As a result, we found that these accretion flows have two types of solutions. 
The first type is an X-ray faint solution characterized by a fast wind velocity (fast-faint solution). 
Another solution denotes an X-ray bright accretion flow associated with a slow wind velocity (slow-bright solution). 
%They are the fast-faint accretion mode and the slow-bright mode.
The fast-faint accretion mode is a steady solution, while the slow-bright solution is unsteady mode. 
Instead, the slow-bright solution is a dividing ridge of the X-ray luminosity evolutions. 
That is, systems fainter than this slow-bright sequence become further fainter until arriving at the fast-faint sequence. 
On the other hand, systems brighter than the slow-bright sequence get brighter. 
However, when the X-ray luminosity becomes sufficiently high, X-ray photoionization effect reduces CAK mechanism which drives stellar wind. 
Hence, the wind flow would be stagnated. 
This stagnation limit of X-ray luminosity gives the upper limit of the X-ray luminosity in wind-fed HMXBs. 
At the same time, persistent faint X-ray sources powered by wind accretions, such as SFXT in quiescence, may follow the fast-faint sequence. 
The parameter space where steady solution can exist is small for systems with short $P_{\rm{orb}}$ and/or small $\dot{M}$. 
Hence, in these systems, the X-ray luminosities of wind-fed X-ray binaries would tend to be settled by the stagnation limit.

Recently, observed examples of faint HMXBs are increasing. 
The existence of the faint persistent sources in wind-fed systems would be a collateral evidence of the fast-faint mode of wind accretion. 
However, in the actual binary systems, one dimensional computation may be insufficient.
%, since in these systems the orbital motion cannot be neglected. 
In the near future, two or three dimensional modeling including orbital motion would figure out the true nature of HMXBs.

%%%%%%%%%%%%%%%%%%%%%%%%%%%%%%%%%%%%%%%%%%%%%%%%%%

\bigskip

%Acknowledgement should be placed at end of main text.
%(NOT after the Appendix.)

We thank an anonymous referee for valuable comments.

%%%
% See the manual for the detail.
%%%

\end{document}